\documentclass{appolb}
\usepackage{epsfig}

\begin{document}
\date{October 9, 2009}
\pagestyle{plain}
\newcount\eLiNe\eLiNe=\inputlineno\advance\eLiNe by -1
\title{Charged Particles are Prevented from Going Faster than the Speed of Light by Light Itself: A Biophysical Cell Biologist's Contribution to Physics
\thanks{Send any remarks to {\tt row1@cornell.edu}}%
}
\author{Randy Wayne
\address{Department of Plant Biology,\\ Cornell University\\Ithaca, New York, 14853 USA\\}}
\maketitle

\begin{abstract}
Investigations of living organisms have led biologists and physicians to introduce fundamental concepts, including Brownian motion, the First Law of Thermodynamics, Poiseuille's Law of fluid flow, and Fick's Law of diffusion into physics. Given the prominence of viscous forces within and around cells and the experience of identifying and quantifying such resistive forces, biophysical cell biologists have an unique perspective in discovering the viscous forces that cause moving particles to respond to an applied force in a nonlinear manner. Using my experience as a biophysical cell biologist, I show that in any space consisting of a photon gas with a temperature above absolute zero, Doppler-shifted photons exert a velocity-dependent viscous force on moving charged particles. This viscous force prevents charged particles from exceeding the speed of light. Consequently, light itself prevents charged particles from moving faster than the speed of light. This interpretation provides a testable alternative to the interpretation provided by the Special Theory of Relativity, which contends that particles are prevented from exceeding the speed of light as a result of the relativity of time.
\end{abstract}
\section{Introduction}

\begin{flushleft}
\textit{``Ask not what physics can do for biology, ask what biology can do for physics.''}
\end{flushleft}
                                                         
\begin{flushright}
-Stanislaw Ulam \cite{1}
\end{flushright}

\par
Cells and the organelles within them live in a world whose dimensions fall between those claimed by the world of macroscopic theoretical physics and those claimed by the world of microscopic theoretical physics \cite{2,3}. In working in this world of neglected dimensions, biophysical cell biologists have used the known laws of physics to make great strides in understanding the physical basis of life \cite{4,5,6,7,8,9,10,11,12,13,14,15,16,17,18,19,20,21,22,23,24,25,26}. However, in working in the world of cellular dimensions, biophysical cell biologists also have a unique opportunity to contribute to ``new physics'' by looking for physical laws that are capable of encompassing theoretical macrophysics and microphysics. This is not such a wild speculation when one considers that much of what we call ``physics'' comes from the study of living organisms \cite{27}. The wave theory of light as a description of diffraction came from Thomas Young's \cite{28} endeavor to understand vision; the discovery of Brownian motion came from Robert Brown's \cite{29,30} study of pollen and pollination; the First Law of Thermodynamics came from Robert Mayer's \cite{31} observation that the venous blood of people living in the warm climate of Java is redder and thus more oxygenated than the venous blood of people living in the colder German climate; the eponymous Law of laminar flow came from Jean Poiseuille's \cite{32} work on describing the flow of blood; and the eponymous law of diffusion came from Adolf Fick's \cite{33} work on describing transmembrane solute movement in kidneys. 
\par
For over four centuries, Newton's \textit{Philosphiae Naturalis Principia Mathematica} has provided a method for describing the frame of the System of the World, and the three laws of motion described in Book One of the \textit{Principia} have formed the theoretical foundations of terrestrial and celestial mechanics  \cite{34,35}. According to Newton's First Law, \textit{``Every body continues in its state of rest, or of uniform motion in a right line, unless it is compelled to change that state by forces impressed on it.''} While this statement is often called the law of inertia, it is better characterized as the assumption of no friction. That is, while the inertia of the body provides a resistance to any change in motion; the body is considered to be \textit{inert} to the medium though which the body moves so that the medium provides no resistance to the movement of the \textit{inert}ial body.
\par
Once one assumes that friction is negligible, Newton's Second Law follows. According to Newton's Second Law, \textit{``The change of motion \textnormal{[}\textit{dmv}/\textit{dt}\textnormal{]} is proportional to the motive force impressed; and is made in the direction of the right line in which that force is impressed.''}  That is, the time rate of change of momentum of an object is proportional to the motive force applied to the object. Assuming, along with Newton, that the mass of the body is a constant, it is the acceleration of the body and not the velocity that is proportional to the applied motive force. Newton's Second Law (\textit{\textbf{F}} = \textit{m}$\frac{d\textit{\textbf{v}}}{d\textit{t}}$), which serves to describe everything from the falling of an apple to the orbit of the Moon, implies that any particle with a constant and invariant mass (\textit{m}) can be accelerated from rest to any velocity (\textit{\textbf{v}}) in time (\textit{t}) by the application of a large enough constant force (\textit{\textbf{F}}) or the application of a small force for a long enough time. 
\par
In contrast to the Laws given by Newton in Book One of the \textit{Principia}, are the lesser-known propositions given in Book Two that describe the motion of bodies in resisting mediums. While the Laws in Book One can be considered to be simple and elegant descriptions of Platonic or ideal situations that have become the foundation of theoretical macroscopic physics, the lesser-known propositions in Book Two, which Newton wrote in haste, can be considered to describe complex fudge factors that depend on a number of variables, including the density and tenacity of the medium as well as the size and shape of the inertial body \cite{36}. These fudge factors, which give the resistance [viscous force] of the medium, were added or subtracted by Newton in order to deduce the magnitude of the ``absolute force'' applied to that body. The effects of these fudge factors on movement were quantified by George Stokes \cite{37}, who provided the foundation for ``non-Newtonian'' physics. By studying mixtures of beach sand or gelatin, Osborne Reynolds \cite{38} and Herbert Freundlich \cite{39}, respectively realized that resistive media could have an infinite number of viscosities that depended on the velocity of the particle moving through the medium. Media whose resistance increased with velocity were called dilatant and media whose resistance decreased with velocity were called thixotropic. Realizing that the structure and composition of the highly dynamic and ever changing cytoplasmic space determined the characteristics of the motile processes that take place within it, William Seifriz, Nobur$\widehat{\textnormal{o}}$ Kamiya and their colleagues championed the use of experimental techniques to characterize the nonlinear resistance to movement provided by the living protoplasm itself \cite{40,41,42,43,44,45,46,47,48,49}. As I will describe below, by using biophysical cell biological thinking, it is possible to integrate the nonlinear properties of the viscous force into Newton's Second Law to construct a simple, elegant, quantitative and testable physical theory that is more robust than Newton's Second Law in that it also encompasses microphysical phenomena. 
\par
Influenced by the field theory of Maxwell \cite{50,51}, scientists in the late nineteenth century generally thought that the elementary unit of electricity known as an electron was nonmaterial in nature and resulted from the release of a center of strain in the nonmaterial aether \cite{52,53,54,55,56}. Going against the conventional wisdom, J. J. Thomson \cite{57} resurrected Gustav Fechner's and Wilhelm Weber's idea that electrons might be corpuscles of matter rather than a released element of an nonmaterial aether and applied Newton's Laws to determine the mass of the electrons that made up cathode rays \cite{58,59,60}. While Newton's Second Law was known to describe the motion of corpuscular bodies from apples to planets, it failed to describe the motion of the corpuscular carrier of electrical charge. In his characterization of the mass-to-charge ratio of the electron, Thomson \cite{61} found that the amount of force required to accelerate an electron increased nonlinearly as the electron's velocity asymptotically approached the speed of light (c). This indicated to Thomson that the mass of the electron increased with velocity. Thomson \cite{62,63} concluded that the increased mass followed from Maxwell's equations and was a result of the magnetic field produced in the aether by the moving charge. 
\par
Interpreting the mass of an electron did not turn out to be straightforward. Hendrik Lorentz \cite{64,65,66} postulated that the mass of an electron moving through an isotropic space was anisotropic and calculated that the longitudinal mass or the component of the mass of an electron parallel to the direction of motion increased with an increase in velocity by a factor of $\sqrt[3]{1 - v^{2}/c^{2}}$, while the transverse mass or the component of the mass of an electron perpendicular to the direction of motion increased with an increase in velocity by a factor of $\sqrt{1 - v^2/c^2}$. The reciprocal of the latter expression is commonly known as the Lorentz factor ($\gamma$).
\par
Based only on his postulates of the principle of relativity and the constancy of the speed of light, Albert Einstein \cite{67} independently derived the velocity-dependent longitudinal and transverse components of the mass of the electron in the dynamical part of the paper entitled, \textit{On the electrodynamics of moving bodies} and concluded that the increase in mass might result from the withdrawal of energy from the electrostatic field and thus might only be apparent. He further wrote that \textit{``in comparing different theories of the motion of the electron we must proceed very cautiously.''} Nevertheless, for many physicists, the interpretation that the mass of an electron increased in a velocity-dependent manner was useful in explaining the nonlinear nature of the force-acceleration relation for an electron \cite{68,69,70,71,72,73}. 
\par
While various mechanical and electromagnetic interpretations of the relationship between impulse (the product of force and time) and velocity were proffered in the first decades of the 20th century \cite{74,75}, by the 1920s, there was a general agreement that the nonlinear relationship between impulse and velocity was explained in terms of the relativity of space-time \cite{76,77,78,79}. That is, the duration of time is a relative quantity that depends on the relative velocity of the electron and the experimenter in the laboratory frame.
\par
According to this kinematic space-time interpretation of Einstein's Special Theory of Relativity, the duration of time (d\textit{t}) varies in a velocity-dependent manner according to the following equation:
\begin{equation}
dt_{proper} = \sqrt{(1 - v^2/c^2)}\ \ dt_{improper},
\end{equation}
where $dt_{proper}$ and $dt_{improper}$ are the proper duration and improper duration of time, respectively. According to the space-time interpretation of the Special Theory of Relativity, the force-acceleration relation is nonlinear because a moving particle experiences a constant force for a shorter duration of time ($dt_{proper}$) than it would if the particle were at rest in the laboratory frame ($dt_{improper}$).  That is, as a particle goes faster and faster, it experiences the motive force for a shorter and shorter duration of time.
\par
Since movement is one of the fundamental attributes of life \cite{4}, biophysical cell biologists spend a great deal of time investigating movement, including the movement of mechanochemical proteins such as myosin, dynein and kinesin \cite{80,81,82,83}; the movement of water and ions through channels in membranes \cite{84,85,86,87,88,89}; the movement of small molecules through plasmodesmata \cite{90}; the movement of proteins from their site of synthesis to their site of action \cite{91}; the movement of membrane vesicles, tubules \cite{92,93,94}; and organelles \cite{95,96} throughout the viscous cytoplasm \cite{97,98}; the movement of chromosomes during mitosis and meiosis \cite{99,100,101}; the movement of cilia and flagella \cite{102,103}; and the movement of the polymers of the extracellular matrix that allow the growth of plant cells \cite{104,105,106,107}. Each one of these cellular movements, as well as others not listed here, involves a motive force that must overcome a non-negligible viscous force. Consequently, the acceleration is not proportional to the applied force. In fact, given the low Reynolds Numbers \cite{108}, which are system-dependent dimensionless numbers that relate the motive force to the viscous force \cite{109}, in a cell, movement ceases immediately after the removal of a motive force, and it is the velocity rather than the acceleration of a body that is proportional to the motive force \cite{110,111,112}.  
\par
Given the prominence of viscous forces within and around a cell \cite{113,114} and the experience of identifying and quantifying such resistive forces, the first question a biophysical cell biologist asks when confronted with the nonlinear relationship between force and the acceleration of an electron is,  ``Can the nonlinear relationship between force and acceleration observed for particles moving at speeds approaching the speed of light be explained by invoking the presence of a viscous force (without reintroducing the contentious nineteenth century aether)?''  Biophysical cell biologists are at an advantage in discovering the viscous forces that cause moving particles to respond to an impulse in a nonlinear manner (Fig. 1).

\begin{figure}[ht]
\begin{center}
\includegraphics[width=12cm]{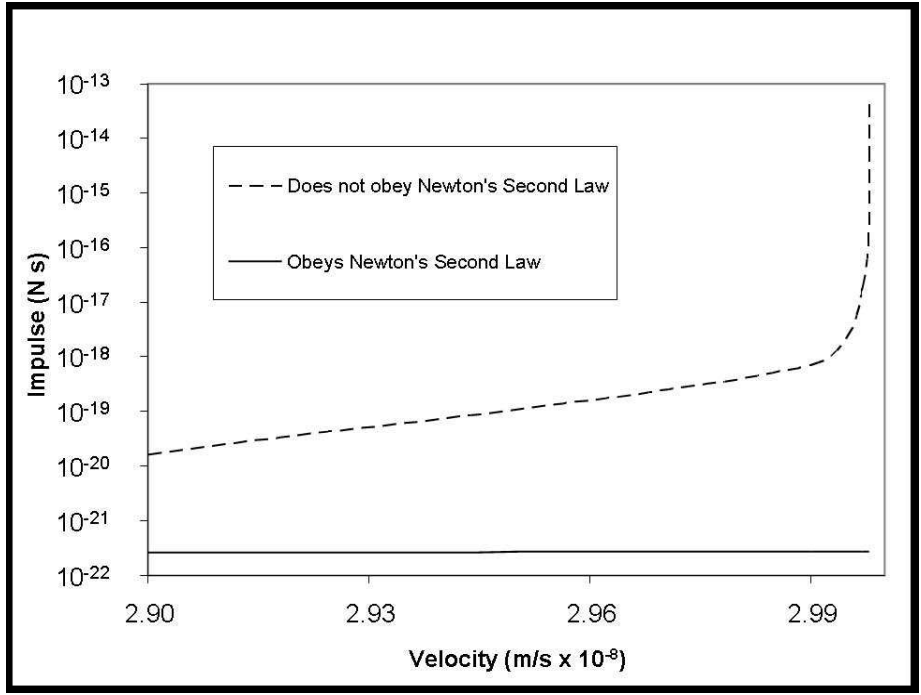}
\end{center}
\caption{The impulse needed to accelerate an electron (\textit{m} = 9.1 x 10$^{-31}$ kg) from rest (\textit{\textbf{v}} = 0) to a given velocity assuming Newton's Second Law is valid and impulse = $\int$ \textit{\textbf{F} dt} = \textit{m\textbf{v}} or that Newton's Second Law is invalid and impulse = $\int$ \textit{\textbf{F} dt}$_{improper}$ $\neq$ \textit{m\bfseries\textbf{v}}.}
\label{f1}
\end{figure}

\section{Results and Discussion}
\par
When a biophysical cell biologist examines the movement of or in a cell, he or she asks, ``What are the physical attributes of the space through which the body moves?'' A biophysical cell biologist strives to answer this question in spite of the fact that the structure and composition or organization of the space itself is not static. Likewise, when one studies the movement of a particle through a space, one must ask the same question. Let us consider the movement of an electron through space where the motive force is provided by an electric field. At any temperature above 0 K, the space consists of a radiation field composed of photons. The photons can be considered to have a black body distribution \cite{115} (Fig. 2).

\begin{figure}[ht]
\begin{center}
\includegraphics[width=12cm]{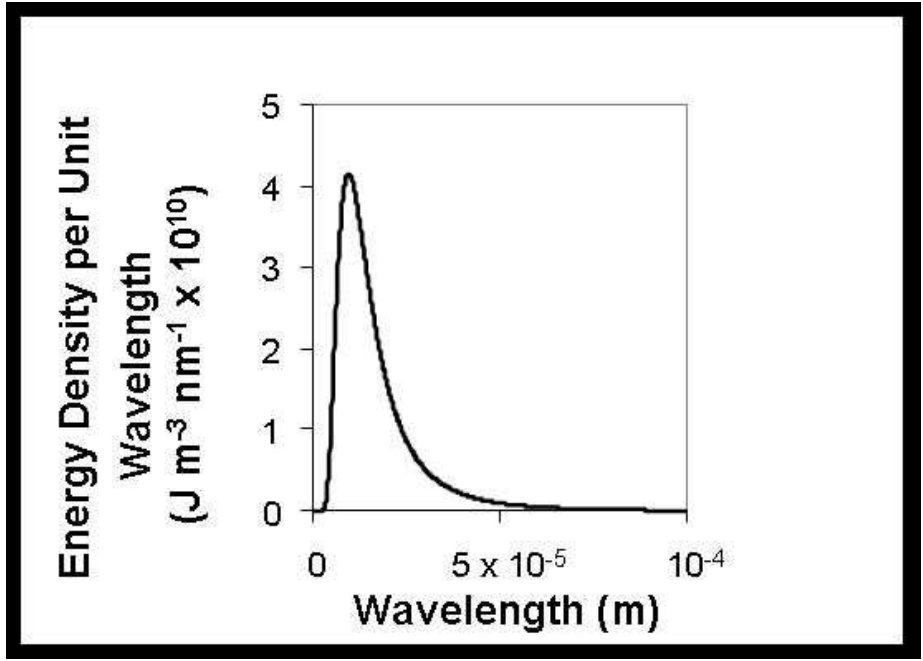}
\end{center}
\caption{The distribution of energy density with respect to wavelength in a photon gas with a temperature of 300 K. The peak will shift to the left and the area under the curve will increase for a photon gas with a temperature $>$ 300 K; and the peak will shift to the right and the area under the curve will diminish for a photon gas with a temperature $<$ 300 K.}
\label{f2}
\end{figure}
\par
The electron moving through the black body radiation field experiences the photons that make up the field as being Doppler shifted \cite{67,116,117}. The photons that collide with the front \cite{118} of the moving electron will be blue shifted (have a higher frequency) and the photons that collide with the back of the moving electron will be red shifted (have a lower frequency) (Fig. 3). 

\begin{figure}[ht]
\begin{center}
\includegraphics[width=12cm]{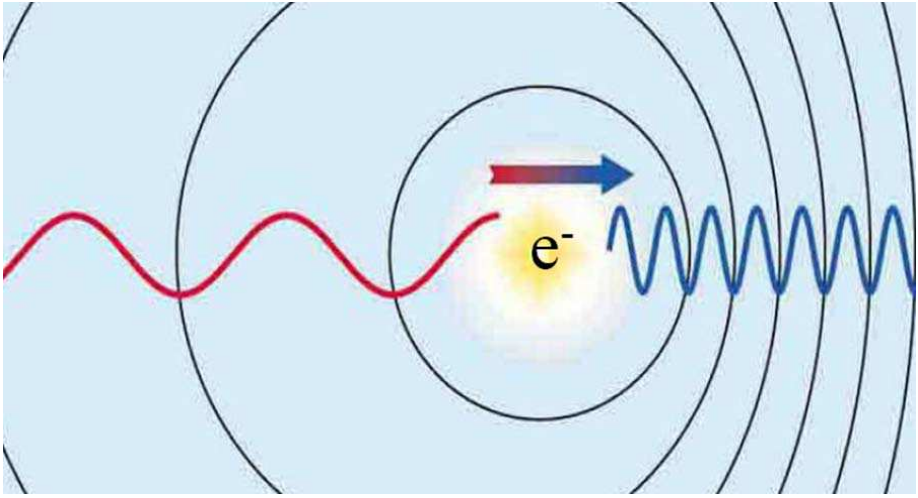}
\end{center}
\caption{An electron moving through a photon gas experiences the photons as being Doppler shifted. The photons that strike the front of the electron and slow it down are blue shifted and the photons that strike the back of an electron and speed it up are red shifted.}
\label{f3}
\end{figure}
\par
Since the electron is moving at a velocity (\textit{\textbf{v}}) relative to the center of momentum of the radiation field, as a result of the Doppler effect, the electron will experience the radiation field as being anisotropic whereas an observer who is at rest with the radiation field will observe it as being isotropic. Consequently, I will describe the radiation experienced by the moving particle with an original relativistic wave equation that describes the propagation of light waves between inertial frames moving relative to each other at velocity \textit{\textbf{v}} and satisfies the requirements set by the Michelson-Morley experiment \cite{119,120,121}. This new relativistic wave equation is given by: 

\begin{equation}
\frac{\partial^2\Psi}{\partial t^2} = \textnormal{c}c'\frac{\sqrt{c - v \cos\theta}}{\sqrt{c + v \cos\theta}}\ \frac{\partial^2\Psi}{\partial x^2},\\ 				
\end{equation}
or
\begin{equation}
\frac{\partial^2\Psi}{\partial t^2} = \textnormal{c}c'\frac{\sqrt{1 - (v \cos\theta)/c}}{\sqrt{1 + (v \cos\theta)/c}}\ \frac{\partial^2\Psi}{\partial x^2},
\end{equation}

where \textit{v} is the absolute value of the velocity and \textit{$\theta$} is the angle between the velocity of a particle and the velocity of a photon (or wave). When the movements of a photon and a particle are parallel, \textit{$\theta$} = 0 radians and cos \textit{$\theta$} = 1; and, when the movements of a photon and a particle are antiparallel, \textit{$\theta$} = $\pi$ radians and cos \textit{$\theta$} = -1 (see appendix 1). 
\par
Different aspects of the speed of a photon or a light wave are represented by c and \textit{c'}. The parameter c, which is absolute and independent of the velocity of the source or the observer, gives the speed of the photon or wave through the vacuum and is equal to the square root of the reciprocal of the product of the electric permittivity ($\epsilon_{o}$) and the magnetic permeability ($\mu_{o}$) of the vacuum. It was the idea of the absolute nature of the speed of light that originally attracted Max Planck to Einstein's Special Theory of Relativity \cite{122}. However, as a consequence of the Doppler Effect, light also has another characteristic speed \textit{c'}, which is local and depends on the relative velocity of the source and observer. \textit{c'} gives the ratio of the angular frequency ($\omega$) of the source in its inertial frame to the angular wave number (\textit{k}) observed in any inertial frame (\textit{c'} $\equiv$ $\omega_{source}/\textit{k}_{observer}$). At any relative velocity between the source and the observer, c\textit{c'}($\frac{\sqrt{1 - (\textit{v} cos \theta)/c}}{\sqrt{1 + (\textit{v} cos \theta)/c}}) = c^{2}$. After canceling c on both sides, we get a relativistic dispersion relation that must be satisfied when the general plane wave solution of the relativistic wave equation has the form: $\Psi = \Psi_{o} e^{i(k_{observer} r - \omega_{source}\frac{\sqrt{c - v cos\theta}}{\sqrt{c + v cos \theta}}t)}$. The relativistic dispersion relation is:
\begin{equation}
\textit{c'}(\frac{\sqrt{1 - (\textit{v} \cos \theta)/c}}{\sqrt{1 + (\textit{v} \cos \theta)/c}}) = c
\end{equation}
At \textit{v} = 0, the new relativistic wave equation reduces to Maxwell's wave equation \cite{50}.
\par
Introducing the perspicuous correction factor (c = $\textit{c'}\frac{\sqrt{1 - (\textit{v} \cos \theta)/c}}{\sqrt{1 + (\textit{v} \cos \theta)/c}}$) into Maxwell's wave equation ensures the invariance of this new relativistic wave equation when describing waves traveling between two inertial frames. A relativistic Doppler equation is obtained naturally from the dispersion relation:

\begin{equation}
k_{observer} = k_{source}\frac{\sqrt{1 - (v \cos\theta)/c}}{\sqrt{1 + (v \cos\theta)/c}} \\
\end{equation}
\begin{equation}
k_{observer} = k_{source}\frac{(1 - v \cos\theta)/c)}{\sqrt{1 - (v^2 \cos^2\theta)/c^2}}. 
\end{equation}
The relativistic Doppler equation allows one to transform angular wave number instead of length and duration between inertial frames.
\par
According to eq. (6), the angular wave numbers in inertial frames moving relative to each other at velocity \textbf{\textit{v}} are related to each other by a Galilean transformation in the numerator and a Lorentz-like transformation in the denominator. The Galilean transformation in the numerator is dominant at low velocities while the Lorentz-like transformation in the denominator is dominant at high velocities.
\par
The experimental observations of Ives and Stillwell \cite{117} on the effect of velocity on the displacement of the spectral lines of hydrogen confirm the utility and validity of using the new relativistic wave equation. However, whereas the Special Theory of Relativity \cite{67,123} predicts the transverse Doppler Effect in an inertial system, the relativistic wave equation given above does not predict any Doppler shift exactly perpendicular to the velocity of an inertial particle. Both the Special Theory of Relativity and the relativistic wave equation presented above predict that averaging the forward and backward longitudinal Doppler-shifted light will give the Lorentz factor also known as the ``time dilation'' factor as observed by Ives and Stillwell \cite{117}. Consequently, experiments that average the forward and backward longitudinal Doppler shifts \cite{117,124} are consistent with both treatments. 
\par
The linear momentum of a photon is given by $\hbar$\textit{k}, where $\hbar$ is Planck's constant divided by 2$\pi$, and \textit{k}, the angular wave number, is equal to 2$\pi$/$\lambda$ \cite{125}. As a consequence of the relativistic Doppler effect, upon interacting with a photon, the change in the linear momentum of an electron (d$\hbar$\textit{k}$_{\textit{electron}}$) moving with speed \textit{v} is velocity dependent and is given by:

\begin{equation}
\textnormal{d}\hbar k_{electron} = \hbar k_{source}\frac{\sqrt{c - v \cos\theta}}{\sqrt{c + v \cos\theta}} \\
\end{equation}
\begin{equation}
\textnormal{d}\hbar k_{electron} = \hbar k_{source} \frac{(1 - (v \cos\theta)/c)}{\sqrt{1 - (v^2 \cos^2\theta)/c^2}}.
\end{equation}

\par
For convenience, I will split eq. (8) into two equations--one for an electron moving parallel (cos $\theta$ = 1) relative to the photons or waves propagating from the source and one for an electron moving antiparallel (cos $\theta$ = -1) relative to the photons or waves propagating from the source. The momentum of the light experienced by the back of an electron traveling parallel to a photon propagating from the source is:

\begin{equation}
\textnormal{d}\hbar k_{electron} = \hbar k_{source}\frac{(1 - v/c)}{\sqrt{1 - v^2/c^2}}.
\end{equation}

\par
This equation is reminiscent of the equation that describes the Compton Effect, where $\hbar$\textit{k}$_{\textit{source}}$ \cite{126} is the momentum of a photon before a collision and d$\hbar$\textit{k}$_{\textit{electron}}$ is the momentum transferred from the photon to the ``back'' of the electron as it is pushed forward \cite{127}. The momentum of the light experienced by a particle traveling antiparallel to light propagating from the source is:

\begin{equation}
\textnormal{d}\hbar k_{electron} = \hbar k_{source}\frac{(1 + v/c)}{\sqrt{1 - v^2/c^2}}.
\end{equation}

\par
This equation is similar to the equation that describes the inverse Compton Effect, where $\hbar$\textit{k}$_{\textit{source}}$ is the momentum of a photon before a collision and d$\hbar$\textit{k}$_{\textit{electron}}$ is the momentum transferred from the photon to the ``front'' of the electron as it is pushed back \cite{128}.
\par
Let us assume that the moving electron interacts with one photon from the front and one photon from the back. The net momentum experienced by this electron moving through a photon gas is antiparallel to the velocity of the electron and would be:

\begin{equation}
\textnormal{d}\hbar k_{electron} = \hbar k_{source}\frac{(1 - v/c) - (1 + v/c)}{\sqrt{1 - v^2/c^2}} \\ 
\end{equation}
\begin{equation}
\textnormal{d}\hbar k_{electron} = \hbar k_{source}\frac{(-\frac{2v}{c})}{\sqrt{1 - v^2/c^2}}.
\end{equation}

\par        					 		
Since the momentum of a blue-shifted photon, striking the front of a moving electron is greater than the momentum of a red-shifted photon striking the back of a moving electron, the momentum of the moving electron decreases. The average decrease of momentum (\textit{p} = \textit{mv} = d$\hbar$\textit{k}$_{\textit{electron}}$) experienced by a moving electron upon colliding with one coaxial photon in a photon gas would be \cite{129}:

\begin{equation}
\textnormal{d}\hbar k_{electron} = -\frac{(\frac{1}{2})\hbar k_{source}(\frac {2v}{c})}{\sqrt{1 - v^2/c^2}} \\ 
\end{equation}
\begin{equation}
\textnormal{d}\hbar k_{electron} = -\frac{\hbar k_{source}(\frac{v}{c})}{\sqrt{1 - v^2/c^2}},	  	
\end{equation}

where the negative sign indicates that the momentum of the electron moving through the radiation field decreases. 
\par
Since the ``average photon'' in the photon gas can strike the moving electron at any angle from 0 to $\pm \frac{\pi}{2}$ with differing effectiveness, the average transfer of momentum \cite{130} from the radiation field to the electron is $\hbar$\textit{k}$_{\textit{source}}$$\frac{(\frac{1}{4})(\frac{v}{c})}{\sqrt{1 - v^2/c^2}}$ for a single collision. As a consequence of the omnipresent existence of a photon gas in space at all temperatures greater than absolute zero, contrary to Newton's First Law, moving charged particles are not inert to the photons in the photon gas through which they move and all bodies composed of charged particles will slow down. In an adiabatic photon gas, this will result in an increase in temperature and an increase in the peak angular wave number of the photons in the photon gas. In an isothermal photon gas, this will result in an expansion of the photon gas. Since a photon gas exists at all temperatures above absolute zero, Newton's First Law is only valid for charged particles at absolute zero and absolute zero, according to the Third Law of Thermodynamics, is unattainable \cite{131}. By taking into consideration the thermodynamics of the photon gas occupying the space through which the electrons move, I will recast Newton's Second Law in an alternative form that applies to charged particles \cite{132} moving at velocities close to the speed of light. 
\par
The velocity-dependent relativistic Doppler-shifted momentum of the photon gas provides the basis for a velocity-dependent viscous force (\textbf{F}$_{Dopp}$) that counteracts the applied force (\textbf{F}$_{app}$)  used to accelerate a particle. 

\begin{equation}
\textbf{F}_{app} + \textbf{F}_{Dopp} = m\frac{d\textbf{v}}{dt},
\end{equation}

where \textbf{F}$_{app}$ and \textbf{F}$_{Dopp}$ are antiparallel by definition.
\par
The viscous force exerted by the radiation field on a moving particle is a function of the collision rate between the moving particle and the photons in the field. The collision rate ($\frac{\textit{dn}}{\textit{dt}}$) depends on the photon density ($\rho$), the speed of the particle (\textit{v}) and the cross section of the photon ($\sigma$) according to the following equation:

\begin{equation}
\frac{dn}{dt} = \rho v \sigma.
\end{equation}

\par
The photon density is a function of the absolute temperature. The absolute temperature on Earth and in the cavity of some accelerators, including the LINAC at Stanford University, is close to 300 K, while the absolute temperature of the cosmic microwave background radiation and in the cavity of other accelerators, including the LINAC at Jefferson Laboratory and the Large Hadron Collider at CERN, is 2.73 K. 
\par
Assuming a blackbody distribution of energy, the photon density can be calculated from Planck's black body radiation distribution formula. According to Planck \cite{115}, the energy density per unit wavelength interval (\textit{u}) is given by:

\begin{equation}
u = \frac{8\pi hc}{\lambda^5}\frac{1}{exp[\frac{hc}{\lambda kT}] - 1}.
\end{equation}

\par	
The peak wavelength of a photon gas can be obtained by differentiating eq. (17) with respect to wavelength or by simply using Wien's distribution law: 

\begin{equation}
\lambda_{peak} = 2.89784\ \textnormal{x}\ 10^{-3} \textnormal{mK}/T = w/T,
\end{equation} 

where \textit{w} is called the Wien coefficient and is equal to 2.89784 x 10$^{-3}$ mK. The peak wavelengths ($\lambda_{peak}$) in 300 K and 2.73 K radiation fields are 9.66 x 10$^{-6}$ m and 1.87 x 10$^{-3}$ m, respectively. The energies of photons with these wavelengths are given by Planck's equation:

\begin{equation}
E = \frac{hc}{\lambda_{peak}},
\end{equation}

and are 2.06 x 10$^{-20}$ J/photon and 1.06 x 10$^{-22}$ J/photon for the peak photons in a 300 K and 2.73 K radiation field, respectively. The total energy density (\textit{U}) of a radiation field can be determined by using the Stefan-Boltzmann equation or by integrating eq. (17) over wavelengths from zero to infinity \cite{133}.

\begin{equation}
U = \int {u\ d\lambda}, \\ 
\end{equation}
\begin{equation}
U = \int {\frac{8\pi hc}{\lambda^5}\frac{1}{(exp[\frac{hc}{\lambda kT}] - 1)}\ d\lambda}, \\ 
\end{equation}
\begin{equation}
U = \frac{8\pi k^4T^4}{c^3 h^3} \int {\frac{x^3}{[exp(x) -1]}\ dx}, \\ 
\end{equation}
\begin{equation}
U = \frac{8\pi k^4T^4}{c^3h^3}(\frac{\pi^4}{15}), \\ 
\end{equation}
\begin{equation}
U = \frac{8\pi^5 k^4}{15c^3h^3}\ (T^4) = 7.57\ \textnormal{x}\ 10^{-16}\ (T^4).
\end{equation}  

\par
The quantity 7.57 x 10$^{-16}$ J m$^{-3}$ K$^{-4}$, known as the radiation constant, is equal to (4$\sigma_{B}$/c), where $\sigma_{B}$ represents the Stefan-Boltzmann constant (5.6704 x 10$^{-8}$ J m$^{-2}$ s$^{-1}$ K$^{-4}$). The total energy densities (\textit{U}) of photon gases with temperatures of 300 K and 2.73 K are 6.13 x 10$^{-6}$ J/m$^{3}$ and 4.02 x 10$^{-14}$ J/m$^{3}$, respectively. The photon densities ($\rho$) in 300 K and 2.73 K radiation fields, which are obtained by dividing eq. (24) by eq. (19), are 2.98 x 10$^{14}$ photons/m$^{3}$ and 3.79 x 10$^{8}$ photons/m$^{3}$, respectively.
\par
Although light is often modeled as an infinite plane wave or a mathematical point, the phenomena of diffraction and interference indicate that a photon has neither an infinite nor a nonexistent width, but something in between \cite{134,135}. By analyzing the relationship between the size of silver halide grains and the light-induced darkening of film, Ludwik Silberstein introduced the cross sectional area of Einstein's light quantum or \textit{light dart} as a useful working hypothesis to quantitatively describe the photographic process \cite{136,137,138}. Silberstein calculated the cross sectional area of a 470 nm photon to be between 8.1 x 10$^{-15}$ m$^{2}$ and 97.3 x 10$^{-15}$ m$^{2}$. I also assume that a photon has a finite width as well as a wavelength and that its geometrical cross section \cite{139} ($\sigma$) is given by:

\begin{equation}
\sigma = \pi r^2,
\end{equation}

where \textit{r} is the radius of the photon (see appendix 2). The radius of a photon can be estimated by following the example of Niels Bohr and using a mixture of classical and quantum reasoning. I make use of the fact that all photons, independent of their wavelength, have the same quantized angular momentum (\textit{L} = $\hbar$), and that classically, angular momentum is equal to \textit{m$\omega$$\Gamma$r$^{2}$} or $\omega$\textit{I}. \textit{I} is the moment of inertia and $\Gamma$ is a geometrical factor that relates \textit{I} to \textit{mr$^{2}$} such that $\Gamma$ = \textit{I}/\textit{mr$^{2}$}. I am assuming that $\Gamma$ is equal to unity, which would be correct if the photon consisted of a point mass at the end of a mass-less string of length \textit{r}. 
\par
By using \textit{E} = \textit{m}c$^{2}$ = $\hbar\omega$ and assuming that the equivalent mass (\textit{m}) of a photon that interacts with matter is given by $\hbar\omega$/c$^{2}$, then the radius (\textit{r}) of a photon will be equal to $\sqrt{\frac{\hbar}{m\omega}}$ = $\sqrt{\frac{\hbar c^{2}}{\hbar\omega^{2}}}$ = $\sqrt{\frac{c^{2}}{\omega^{2}}}$ = c/$\omega$ = $\frac{1}{k}$ = $\frac{\lambda}{2\pi}$, which is the reciprocal of the angular wave number \cite{140} and equal to the wavelength of the photon divided by 2$\pi$. Thus, the geometrical cross section, which is related to its angular wave number and wavelength, is given by:

\begin{equation}
\sigma = \pi (\frac{1}{k})^2 = \pi (\frac{\lambda}{2\pi})^2 = \frac{\lambda^2}{4\pi}.
\end{equation}

\par
According to this reasoning, the cross sections of thermal (300 K) photons and microwave (2.73 K) photons are 7.43 x 10$^{-12}$ m$^{2}$ and 2.78 x 10$^{-7}$ m$^{2}$, respectively. The cross sectional area of 470 nm photons given by eq. (26) is 17.6 x 10$^{-15}$ m$^{2}$, consistent with the values of 470 nm photons given by Silberstein \cite {136,137}.
\par
According to eq. (16), the collision rate ($\frac{\textit{dn}}{\textit{dt}}$) between a moving charged particle and photons in a photon gas is dependent on the speed of the charged particle. After factoring in the photon densities and the cross section of the peak photons, I find that the collision rates are equal to (2214.14 collisions/m)\textit{v} and (105.36 collisions/m)\textit{v }for 300K and 2.73 K radiation fields, respectively, where \textit{v} is the speed of the charged particle relative to the observer who experiences the photon gas as being isotropic. For a given speed, the collision rate increases with the temperature of the radiation field. This is because the photon density increases with the third power of the temperature while the geometrical cross section decreases with the first power of the temperature. In a 300 K thermal radiation field, at speeds approaching the speed of light, the collision rate will be about 6.64 x 10$^{11}$ s$^{-1}$, while it will be about 3.16 x 10$^{10}$ s$^{-1}$ for a 2.73 K microwave radiation field. 
\par
The velocity-dependent viscous force (\textbf{F}$_{Dopp}$) exerted on a moving electron by the photons in the photon gas is given by the product of the collision rate and the average velocity-dependent momentum of a photon:

\begin{equation}
\textbf{F}_{Dopp} = -\frac{(\rho \sigma v)(\hbar k_{source})(\frac {1}{4})(\frac {v}{c})}{\sqrt{1 - v^2/c^2}}, \\ 
\end{equation}
\begin{equation}
\textbf{F}_{Dopp} = -\frac{(\rho \sigma h/4\lambda_{source})(\frac{v^2}{c})}{\sqrt{1 - v^2/c^2}}.
\end{equation}

\par
At \textit{v} = 0, there is no net viscous force and the average momenta (h/4\textit{$\lambda_{source}$}) of photons in a thermal radiation field and a microwave radiation field are 1.72 x 10$^{-29}$ kg m/s and 8.86 x 10$^{-32}$ kg m/s, respectively. Since when \textit{v} = 0, the collisions with the photons in the photon gas are random, the electron will exhibit Brownian motion.
\par
We can define the product of \textit{$\rho$} and \textit{$\sigma$} as the linear photon density (\textit{$\rho_{L}$}), replace h in eq. (28) with $\frac{e^{2}}{\epsilon_{o}c\alpha}$, using the definition of the fine structure constant \cite{138,139} ($\alpha$), substitute $\mu_{o}$ for 1/$\epsilon_{o}c^{2}$, and replace \textit{$\lambda_{source}$} with \textit{w}/\textit{T} to get,

\begin{equation} 
F_{Dopp} = -\frac{[\rho_{L} T e^2\mu_{o}/4w\alpha](v^2)}{\sqrt{1 - v^2/c^2}}. 
\end{equation}

\par
This expression of the viscous force shows explicitly that the viscous force depends on the temperature, the square of the charge of the moving particle, the magnetic permeability of the vacuum, and the fine structure constant, which quantifies the strength of the interaction between a charged particle and the radiation field. The viscous force vanishes, as either the charge of the moving particle or the temperature goes to zero. 
\par
The linear photon density ($\rho_{L}$, in m$^{-1}$) is only a function of temperature and can be written as $\frac{\sigma_{B}w^{3} T}{\pi hc^2}$ or $7.375\ $m$^{-1}$ K$^{-1}$ $T$. By combining the constants in eq. 29, the viscous force experienced by a univalent particle can be expressed exclusively in terms of temperature and velocity:

\begin{equation} 
\textbf{F}_{Dopp} = -[1.41\ \textnormal{x}\ 10^{-39}\ \textnormal{N}\ \textnormal{s}^{2}\ \textnormal{m}^{-2}\  \textnormal{K}^{-2}]\ T^{2}\frac{(v^2)}{\sqrt{1 - v^2/c^2}}. 
\end{equation}

\par
The magnitude of the friction on a univalent particle caused by the viscous force depends on the velocity of the particle and the square of the temperature. The coefficient of friction of the photon gas is given by $\frac{F_{Dopp}}{v}$ = [1.41 x 10$^{-39}$ $\textnormal{N}\ \textnormal{s}^{2}\ \textnormal{m}^{-2}\ \textnormal{K}^{-2}]\  T^{2} \frac{(v)}{\sqrt{1 - v^2/c^2}}$. The power dissipated by the photon gas is given by $vF_{Dopp}$. The power dissipated not only depends on the temperature but will increase the temperature and/or the volume of the space composed of the photon gas. Such an effect may have been important in the expansion of the Universe.
\par
The equation of motion that accounts for the temperature and velocity-dependent viscous force experienced by a univalent particle is:
\begin{equation}
F_{app} - [1.41\ \textnormal{x}\ 10^{-39}\ \textnormal{N}\ \textnormal{s}^{2}\ \textnormal{m}^{-2}\  \textnormal{K}^{-2}]\  T^{2}\frac{(v^2)}{\sqrt{1 - v^2/c^2}} = m\frac{dv}{dt}
\end{equation}
which is equivalent to:
\begin {equation}
F_{app} + F_{Dopp} = m\frac{dv}{dt}.
\end{equation}

\par
Any equation that describes motion at velocities close to the speed of light must reduce to Newton's Second Law for motion at small velocities \cite{143}. Indeed eq. (31) reduces to Newton's Second Law when $v <<$ c and $F_{app}$ $>>$ [1.41 x 10$^{-39}$ $\textnormal{N}\ \textnormal{s}^{2}\ \textnormal{m}^{-2}\ \textnormal{K}^{-2}]\  T^{2}\frac{(v^2)}{\sqrt{1 - v^2/c^2}}$. The integration of eq. (31) with respect to distance yields a quantitative relationship between the apparent mass of a charged particle and energy \cite{144}. 
\par
At temperatures greater than 0 K, the photons, which make up the photon gases that occupy all space, act as dilatant, shear-thickened, viscous ``non-Newtonian'' solutions. While I am not attempting to resurrect the complex and self-contradictory 19th century aether by any means, oddly enough, the rheological concept of dilatancy was first developed by Osborne Reynolds \cite{38,145} when he was contemplating the nature of the luminous ether. Quantum electrodynamics handles the viscous force in terms of renormalization (see appendix 3). 
\par
Equation (31), which is based on biophysical cell biological analogies \cite{146} contrasts with Planck's \cite{147} relativistic version of Newton's Second Law, which is:

\begin{equation}
F_{app} = \frac {d}{dt}\frac{mv}{\sqrt{1 - v^2/c^2}}.
\end{equation}

\par
Depending on the assumptions \cite{148,149,150,151,152,153} used to differentiate eq. (33), the following solutions have been proffered:

\begin{equation}
F_{app}\sqrt{1 - v^2/c^2} = m(\frac {dv}{dt}),\\
\end{equation}

and

\begin{equation}
F_{app}\sqrt[3]{1 - v^2/c^2} = m(\frac {dv}{dt}).
\end{equation}

\par
In contrast to eq. (31), where the term that includes the ``Lorentz factor'' is subtracted from the applied force, in eqs. (34) and (35), the applied force is multiplied by the terms that include the ``Lorentz factor.'' Equations (31), (34), and (35) all predict that the relation between the impulse needed to accelerate an electron from rest to a given velocity and that velocity will be nonlinear (Fig. 4). However, the opto-mechanical Doppler Effect model based on biophysical cell biological reasoning, further predicts that the applied force necessary to overcome the viscous force in order to accelerate a charged particle from rest to velocity $v$ in time $t$ will be temperature dependent. 

\begin{figure}[ht]
\begin{center}
\includegraphics[width=12cm]{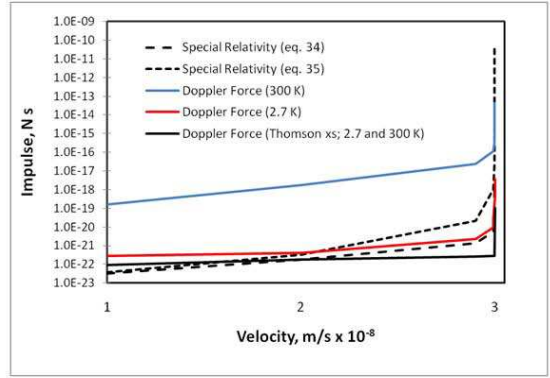}
\end{center}
\caption{The impulse needed to accelerate an electron from rest to a given velocity at two different temperatures as predicted by the opto-mechanical Doppler model (eq. 31) using the geometrical cross section, two interpretations of the Special Theory of Relativity (eqs. 34 and 35), and the opto-mechanical model using the Thompson scattering cross section (see Appendix 2). Equations (34) and (35) have no temperature dependence.}
\label{f4}
\end{figure}
\par
A test of whether the magnitude of the impulse needed to accelerate an electron from rest to a given velocity is influenced by temperature will determine whether the limiting speed of light is a consequence of the viscous force provided by the Doppler-shifted photons that populate every space as predicted by the biophysical cell biologist's model or due to the relativity of time as predicted by Einstein's Special Theory of Relativity. The test can be performed by measuring the impulse necessary to accelerate an electron from rest to a given velocity \cite{154}  at different temperatures.  
\section{Conclusion}
\par
While it is possible that there is more than one mechanism to prevent charged particles from moving faster than the speed of light, the causal, picturesque and testable biophysical cell biologist's opto-mechanical working hypothesis presented here indicates that light can act as an ultimate speed limit to any charged particle because, at velocities approaching the speed of light, the photon gas that occupies the space through which the particles move is not a ``Newtonian'' photon gas with a single and trivial viscosity but is a ``non-Newtonian'', shear-thickened or dilatant photon gas that becomes infinitely viscous as the velocity of a moving charged particle approaches the speed of light. That is, as a result of the Doppler Effect, light itself, and not the relativity of time, may prevent charged particles from moving faster than the speed of light.
\section{Acknowledgement}
\par
I dedicate this paper to my brother Scott Wayne.
\section{Appendix 1}
\par
It is possible that Maxwell's second order wave equation \cite{50} is not the best starting point for describing relativistic phenomenon since his second order wave equation does not contain a relative velocity term. Consequently I introduced a new form-invariant relativistic wave equation, which is consistant with known phenomena that depend on the relative velocity of the source and observer. Since the derivations of eqs. (31) and (32) critically depend on the introduction of this new relativistic wave equation, here I show that the new relativistic wave equation is consistent with the two fundamental principles upon which Einstein based the Special Theory of Relativity \cite{67}. I assume that the new relativistic wave equation, which is form invariant and consistent with the principle of relativity, is the equation of motion that describes the properties of light experienced by an electron moving at velocity $v$ relative to the center of momentum of the photon gas, which is at rest relative to the inertial frame of the observer:
\begin{equation}
\frac{\partial^2\Psi}{\partial t^2} = \textnormal{c}c'\ \frac{\sqrt{\textnormal{c} - v \cos\theta}}{\sqrt{\textnormal{c} + v \cos\theta}}\ \nabla^2\Psi.
\end {equation}
\par
I also assume that the following equation is a general plane wave solution to the second order relativistic wave equation given above: 
\begin{equation}
\Psi = \Psi_{o} e^{i(\textbf{k}_{observer}\cdot\textbf{r}\  -\  \omega_{source}\frac{\sqrt{c - v cos\theta}}{\sqrt{c + v cos \theta}}t)}.
\end{equation}
\par
By substituting eq. (37) into eq. (36) and taking the spatial and temporal partial derivatives, we can obtain the relativistic dispersion relation:
\begin {equation}
\textnormal{c}c' (\frac{\sqrt{\textnormal{c} - v \cos\theta}}{\sqrt{\textnormal{c} + v \cos\theta}}) i^{2}k_{observer}^{2}\Psi = i^{2}\omega_{source}^{2}(\frac{\textnormal{c} - v \cos\theta}{\textnormal{c} + v \cos\theta})\Psi.
\end{equation}
\par
After canceling like terms, we get:
\begin{equation}
\textnormal{c}c'k_{observer}^{2} = \omega_{source}^{2}\frac{\sqrt{\textnormal{c} - v \cos\theta}}{\sqrt{\textnormal{c} + v \cos\theta}}.
\end{equation}
\par
Since $c'$ $\equiv$ $\frac{\omega_{source}}{k_{observer}}$, the above equation simplifies to:
\begin {equation}
\textnormal{c}k_{observer} = \omega_{source}\frac{\sqrt{\textnormal{c} - v \cos\theta}}{\sqrt{\textnormal{c} + v \cos\theta}}.
\end{equation}
\par
By solving for c, the speed of the wave, we get the relativistic dispersion relation:
\begin {equation}
\textnormal{c} = \frac{\omega_{source}}{k_{observer}}\frac{\sqrt{\textnormal{c} - v \cos\theta}}{\sqrt{\textnormal{c} + v \cos\theta}}.
\end{equation}
\par
The relativistic dispersion relation tells us that while the observed angular wave number of light and consequently its observed momentum varies in a velocity-dependent manner, the speed of light (c) is invariant and independent of the motion of the source or observer, consistent with the principle of the constancy of the speed of light. That is, while the speed of light is independent of the relative velocity between the source and the observer, the relative velocity between the source and the observer results in a ``stretching'' or ``compressing'' of the light wave without changing its speed. Since the stretched light waves transfer a smaller quantity of momentum per unit time compared with the compressed light waves, an electron moving through a photon gas with a temperature greater than 0 K experiences a velocity-dependent viscous force.  
\section{Appendix 2}
\par
The degree of nonlinearity between the applied force and the rate of change in momentum predicted by eqs. (31) and (32) depends quantitatively on the calculated geometrical cross sections of the photons that comprise the photon gas through which the charged particle moves. It is natural to wonder what is the effect of replacing the geometrical cross section of a photon with the Thomson scattering cross section \cite{155}, which is not a geometrical cross section but a measure of the probability that a photon will interact with an electron. The Thomson cross section is used to model the Sunyaev-Zel'dovich effect in which high energy electrons are decelerated by the cosmic microwave background and produce x-rays as a result of inverse Compton scattering  \cite{156}. In contrast to the geometrical cross section, the Thomson scattering cross section is wavelength- and thus temperature-independent. The Thomson scattering cross section ($\sigma_{T}$) is 6.6524586 x 10$^{-29}$ m$^{2}$, which is orders of magnitude smaller than the geometrical cross section of the photon presented in this paper or calculated by Silberstein \cite{136,137}. Using the Thomson scattering cross section instead of the geometrical cross section, the formula for the temperature- and velocity-dependent viscous force given by eq. (31) becomes:
\begin{equation}
F_{app} - \frac{\sigma_{T} \sigma_{B}}{c^{3}} T^{4}\frac{(v^2)}{\sqrt{1 - v^2/c^2}} = m\frac{dv}{dt}
\end{equation}
After combining the constants we get:
\begin{equation}
F_{app} - [1.40\ \textnormal{x}\ 10^{-61}\ \textnormal{N}\ \textnormal{s}^{2}\ \textnormal{m}^{-2}\  \textnormal{K}^{-4}]\  T^{4}\frac{(v^2)}{\sqrt{1 - v^2/c^2}} = m\frac{dv}{dt}
\end{equation}
\par 
While the temperature dependence is greater in the equation that uses the Thompson scattering cross section than in eq. (31), the coefficient in the equation that uses the Thomson scattering cross section is approximately twenty-two orders of magnitude smaller than the coefficient in eq. (31) that uses the geometrical cross section, and, as shown in Fig. 4, the viscous force that results from the relativistic Doppler effect would only be detectable at velocities infinitessimally close to the speed of light.
\par
Since the acceleration of an electron parallel to the electric field of the incident radiation may also be retarded by the photon gas and as a result of the retardation, dissipate energy in a manner that may affect the scattering cross section, I calculated a scattering cross section using the Larmor formula in a manner following Thomson \cite{155} and Purcell \cite{157} except that I substituted $a^{2} = (\frac{eE}{m} + F_{Dopp})^{2}$ for $a^{2} = e^{2}E^{2}/m^{2}$ where $E$ represents the electric field of the incident radiation. The Larmor formula describes the influence of acceleration on the power emitted by a moving charge with mass $m$ and charge $e$. The resulting formula for the scattering cross section ($\sigma_{s}$) is:
\begin{equation}
\sigma_{s} = \frac{e^{4}}{6 \pi \epsilon_{o}^{2} m^{2} c^{4}} + \frac{2F_{Dopp}e^{3}}{6 \pi \epsilon_{o}^{2} m^{2} c^{4} E} + \frac{F_{Dopp}^{2}e^{2}}{6 \pi \epsilon_{o}^{2} m^{2} c^{4} E^{2}}
\end{equation}
\par
The first term is equal to the Thomson scattering cross section, which is both temperature- and velocity-independent. Thus the temperature- and velocity-dependent cross section can be written as:
\begin{equation}
\sigma_{s} = \sigma_{T} + \frac{2F_{Dopp}e^{3}}{6 \pi \epsilon_{o}^{2} m^{2} c^{4} E} + \frac{F_{Dopp}^{2}e^{2}}{6 \pi \epsilon_{o}^{2} m^{2} c^{4} E^{2}}
\end{equation}
\par
When either the temperature or velocity approaches zero, the above scattering cross section reduces to the Thomson scattering cross section. While the temperature- and velocity-dependent cross section given above is greater than $\sigma_{T}$, it is still small enough that, if it were an accurate representation of the cross section of a photon, then the viscous force caused by the relativistic Doppler effect would only be detectable at velocities infinitessimally close to the speed of light. 
\par
The derivation of the Thomson scattering cross section assumes that the electric field produced by the electron is symmetric. However, the symmetry of the electric field produced by a moving electron may depend on its velocity. For example, the measured value of the electric field, as well as its divergence, may be greater in front of a moving charge than behind it if the measurements occur over a finite time. I postulate that at a distance $r$, the electric field produced by a moving charge ($q_{m}$) may be given by:

\begin{equation}
\textbf{E} = \frac{1}{4 \pi \epsilon_{o}} \frac {1 - \frac{vcos \theta}{c}}{\sqrt{1-\frac{v^{2} cos^{2} \theta}{c^{2}}}}\frac{q_{m}}{r^{2}}\hat{\textbf{r}}.
\end{equation}
Such an asymmetrical distribution of the electric field along the direction of motion contrasts with the symmetrical velocity-dependent distribution of the electric field postulated by the Special Theory of Relativity \cite{158}. By taking into consideration the possibility of a velocity-dependent asymmetry in the electric field of a moving charge it may be possible to derive a more relevant scattering cross section to model the viscous force.
\par
Standard scattering theory based upon interaction cross sections are unable to account for a detectable temperature dependence in the impulse-velocity curve. Consequently, a discovery of a temperature dependence in an experimentally-obtained impulse-velocity curve would support the validity of using the geometrical cross section as well as reify the proposal that light itself prevents charged particles from moving faster than the speed of light. On the other hand, the demonstration of a temperature independence in an experimentally-obtained impulse-velocity curve would support the assumption that the photon is best modeled as a mathematical point with a geometrically undefined interaction cross section and it is the relativity of time that prevents any particle from moving faster than the speed of light. 

\section{Appendix 3}
\par
I consider the mass of a charged particle moving through space composed of a photon gas to be constant and invariant. However, the effect of Doppler-shifted photons on resisting the movement of an electron moving through a photon gas consisting of real photons can be interpreted in terms of an increase in the effective mass of an electron that occurs as a consequence of the dressing or renormalizing of the electron as it interacts with the virtual photons of the quantum electrodynamical vacuum \cite{159}. The velocity- and temperature-dependent ratio of the apparent mass that results from the viscous force to the constant mass ($m$) is given by the following equation: 
\begin{equation}
\frac{m_{apparent}}{m} = \frac{F_{applied}}{F_{applied} + F_{Doppler}} 
\end{equation}
After rearranging, we get the apparent mass or effective mass of an electron:
\begin{equation}
m_{apparent} = \frac {m}{1 + \frac{F_{Dopp}}{F_{applied}}} 
\end{equation}
where the second term in the denominator, which is velocity- and temperature-dependent, is analogous to the self-energy of the dressed electron.


\begin{thebibliography}{99}

\bibitem{1} H. Frauenfelder, P. G. Wolynes and R. H. Austin, \textit{Rev. Mod. Phys.} \textbf{71}, S419 (1999).
\bibitem{2} W. Ostwald, \textit{An Introduction to Theoretical and Applied Colloid Chemistry. The World of Neglected Dimensions}, John Wiley $\&$ Sons, New York, 1922.
\bibitem{3} A. Frey-Wyssling, \textit{Macromolecules in Cell Structure}, Harvard University Press, Cambridge, 1957.
\bibitem{4} T. H. Huxley, in \textit{Lay Sermons, Addresses, and Reviews }, D. Appleton and Co., New York, 1890, p. 120.
\bibitem{5} J. Tyndall, in \textit{Fragments of Science} Vol. II, D. Appleton and Co., New York, 1898, p. 135.
\bibitem{6} A. V. Hill, \textit{Living Machinery}, Harcourt, Brace and Co., New York, 1927.
\bibitem{7} D. Burns, \textit{An Introduction to Biophysics}, Macmillan, New York, 1929).
\bibitem{8} R. H\"ober, \textit{Physical Chemistry of Cell and Tissues}, Blakiston, Philadelphia, 1945.
\bibitem{9} A. Szent-Gy\"orgyi, \textit{Nature of Life}, Academic Press, New York, 1948.
\bibitem{10} A. Szent-Gy\"orgyi, \textit{Introduction to a Submolecular Biology}, Academic Press, New York, 1960.
\bibitem{11} A. Frey-Wyssling, ed, \textit{Deformation and Flow in Biological Systems}, North-Holland Publishing Co., Amsterdam, 1952.
\bibitem{12} M. Tazawa, \textit{Protoplasma} \textbf{48}, 342 (1957).
\bibitem{13} L. V. Heilbrunn, \textit{The Viscosity of Protoplasm}. Protoplasmatologia, Springer-Verlag, Vienna, 1958.
\bibitem{14} N. Kamiya, \textit{Protoplasmic streaming}. Protoplasmatologia. Bd 8,3a, Springer-Verlag, Vienna, 1959.
\bibitem{15} N. Kamiya, \textit{Annu. Rev. Plant Physiol.} \textbf{11}, 324 (1960).
\bibitem{16} N. Kamiya, \textit{Annu. Rev. Plant Physiol.} \textbf{32}, 205 (1981).
\bibitem{17} N. Kamiya, \textit{Bot. Mag. Tokyo} \textbf{99}, 441 (1986).
\bibitem{18} G. von B\'ek\'esy, \textit{Experiments in Hearing}, McGraw-Hill, New York, 1960.
\bibitem{19} D. M. Needham, \textit{Machina Carnis. The Biochemistry of Muscular Contraction in its Historical Development}, Cambridge University Press, Cambridge, 1971.
\bibitem{20} J. C. Eccles, \textit{The Understanding of the Brain}, McGraw-Hill, New York, 1973.
\bibitem{21} A. F. Huxley, \textit{Reflections on Muscle}, Princeton University Press, Princeton, 1980.
\bibitem{22} R. K. Clayton, \textit{Photosynth. Res.} \textbf{19 }, 207 (1988).
\bibitem{23} K. J. Niklas, \textit{Plant Biomechanics. An Engineering Approach to Plant Form and Function}, University of Chicago Press, Chicago, 1992.
\bibitem{24} H. E. Huxley, \textit{Ann. Rev. Physiol.} \textbf{58}, 1 (1996).
\bibitem{25} G. Feher, \textit{Annu. Rev. Biophys. Biomol. Struct.} \textbf{31}, 1 (2002).
\bibitem{26} P. S. Nobel, \textit{Physicochemical and Environmental Plant Physiology}, Academic Press, San Diego, 2009.
\bibitem{27} R. Wayne, \textit{Plant Cell Biology. From Astronomy to Zoology}, Elsevier Academic Press, Amsterdam, 2009.
\bibitem{28} T. Young, \textit{A Course of Lectures on Natural Philosophy and the Mechanical Arts}, Printed for Joseph Johnson, London, 1807.
\bibitem{29} R. Brown, in \textit{The Miscellaneous Botanical Works of Robert Brown} Vol. I, Robert Hardwicke, London, 1866, p. 463. 
\bibitem{30} R. Brown, in \textit{The Miscellaneous Botanical Works of Robert Brown} Vol. I, Robert Hardwicke, London, 1866, p. 479.
\bibitem{31} J. R. Mayer, in W. R. Grove, H. Helmholtz, J. R. Mayer, M. Faraday, J. Liebig, W. B. Carpenter, and E. L. Youmans, \textit{The Correlation and Conservation of Forces. A Series of Expositions}, D. Appleton and Co., New York, 1868, p. 316.
\bibitem{32} J. L. M. Poiseuille, \textit{Experimental Investigations upon the Flow of Liquids in Tubes of Very Small Diameter} Translated by W. H. Herschel. Rheological Memoirs. E. C. Bingham, ed. Vol. 1, Number 1, Lancaster Press, Inc., Easton, 1940.
\bibitem{33} A. Fick, \textit{Phil. Mag.} \textbf{10}, 30 (1855).
\bibitem{34} F. Cajori, \textit{Sir Isaac Newton's Mathematical Principles of Natural Philosophy and his System of the World}, University of California Press, Berkeley, 1946.
\bibitem{35} H. Bondi, in \textit{Let Newton Be! A New Perspective on his Life and Works} eds. J. Fauvel, R. Flood, M. Shortland, and R. Wilson, Oxford University Press, Oxford, 1988, p. 241.
\bibitem{36} R. S. Westfall, \textit{Science} \textbf{179}, 751 (1973).
\bibitem{37} G. G. Stokes, in \textit{Mathematical and Physical Papers} Vol. III, Cambridge University Press, Cambridge, 1922, p. 1.
\bibitem{38} O. Reynolds, \textit{Phil. Mag. Ser. 5.} \textbf{20}, 469 (1885).
\bibitem{39} H. Freundlich, in \textit{A Symposium on the Structure of Protoplasm} ed. W. Seifriz, Iowa State College Press, Ames, 1942. p. 85.
\bibitem{40} H. Freundlich and W. Seifriz, \textit{Z. phys. Chem.} \textbf{104},  233 (1923).
\bibitem{41} W. Seifriz, \textit{J. Rheology} \textbf{1}, 261 (1930).
\bibitem{42} W. Seifriz, \textit{Protoplasm}, McGraw-Hill, New York, 1936.
\bibitem{43} W. Seifriz and J. Plowe, \textit{J. Rheology} \textbf{2}, 263 (1931).
\bibitem{44} N. Kamiya and W. Seifriz, \textit{Exp. Cell Res.} \textbf{6}, 1 (1954).
\bibitem{45} N. Kamiya, \textit{Protoplasma} \textbf{45}, 513 (1956).
\bibitem{46} N. Kamiya, \textit{Annu. Rev. Plant Physiol. Plant Mol. Biol}. \textbf{40}, 1 (1989).
\bibitem{47} N. Kamiya and K. Kuroda, Proc. IVth Intern. Congr. Rheology. Part 4. Symp. Biorheol., John Wiley, New York, 1965, p. 157.
\bibitem{48} N. Kamiya and K. Kuroda, \textit{Biorheology} \textbf{10}, 179 (1973).
\bibitem{49} M. Tazawa, \textit{Cell Struc. Funct.} \textbf{24}, 55 (1999).
\bibitem{50} J. C. Maxwell, \textit{Phil. Trans. Roy. Soc. Lond.} \textbf{155}, 459 (1865) .
\bibitem{51} J. C. Maxwell, \textit{A Treatise on Electricity and Magnetism}, Dover, New York, 1954.
\bibitem{52} O. Lodge, \textit{Modern Views of Electricity}, Macmillan, London, 1889.
\bibitem{53} G. J. Stoney, \textit{Phil. Mag. Ser. 5.} \textbf{38}, 418 (1894).
\bibitem{54} J. Larmor, \textit{Aether and Matter}, Cambridge University Press, Cambridge, 1900.
\bibitem{55} W. Kaufmann, \textit{The Electrician} November 8, 1901. \textit{48}, 95 (1901).
\bibitem{56} E. Whittaker, \textit{A History of the Theories of Aether and Electricity. The Classical Theories}, Thomas Nelson and Sons, London, 1951.
\bibitem{57} J. J. Thomson, \textit{The Corpuscular Theory of Matter}, Archibald Constable $\&$ Co., London, 1907.
\bibitem{58} A. Bennett, R. Heikes, P. Klemens, A. Maradudin and S. Banigan, \textit{Electrons on the Move}, Walker and Co., New York, 1964.
\bibitem{59} E. A. Davis and I. Falconer,  \textit{J. J. Thomson and the Discovery of the Electron}, Taylor $\&$ Francis, London, 1997.
\bibitem{60} I. Falconer, \textit{Physics Education} \textbf{32}, 226 (1997).
\bibitem{61} J. J. Thomson, \textit{Phil. Mag. Ser 5.} \textbf{44}, 293 (1897).
\bibitem{62} J. J. Thomson, \textit{Phil. Mag. Ser. 5.} \textbf{11}, 229 (1881).
\bibitem{63} J. J. Thomson, \textit{Recollections and Reflections}, Macmillan, New York, 1937.
\bibitem{64} H. A. Lorentz, \textit{Proc. Roy. Netherlands Acad. Arts Sci.} \textbf{6}, 809 (1904).
\bibitem{65} H. A. Lorentz, \textit{Problems of Modern Physics}, Ginn and Co., Boston, 1927.
\bibitem{66} H. A. Lorentz, \textit{The Theory of Electrons}, Dover, New York, 1952, 2nd ed.
\bibitem{67} A. Einstein, in \textit{The Collected Papers of Albert Einstein}. Vol. 2. English Translation, Princeton University Press, Princeton, 1989, p. 140.
\bibitem{68} M. Jammer, \textit{Concepts of Mass in Classical and Modern Physics}, Dover, New York, 1961.
\bibitem{69} M. Born, \textit{Einstein's Theory of Relativity}, Dover, New York, 1962.
\bibitem{70} R. P. Feynman, R. B. Leighton and M. Sands, \textit{The Feynman Lectures on Physics} Vol I, Addison-Wesley, Reading, 1963.
\bibitem{71} H. Bondi, \textit{Relativity and Common Sense. A New Approach to Einstein}, Dover, New York, 1964.
\bibitem{72} D. Bohm, \textit{The Special Theory of Relativity}, W. A. Benjamin, New York, 1965.
\bibitem{73} T. R. Sandlin, \textit{Amer. J. Phys.} \textbf{59}, 1032 (1991).
\bibitem{74} J. T. Cushing, \textit{Amer. J. Phys.} \textbf{49}, 1133 (1981).
\bibitem{75} A. K. Wr$\acute{\textnormal{o}}$blewski, \textit{Acta Physica Polonica B} \textbf{37}, 11 (2006).
\bibitem{76} C. G. Adler, \textit{Amer. J. Phys.} \textbf{55}, 739 (1989).
\bibitem{77} N. D. Mermin, \textit{It's about Time. Understanding Einstein's Relativity}, Princeton University Press, Princeton, 2005.
\bibitem{78} S. Bais, \textit{Very Special Relativity}, Harvard University Press, Cambridge, 2007.
\bibitem{79} L. B. Okun, \textit{Amer. J. Phys.} \textbf{77}, 430 (2009).
\bibitem{80} M. P. Sheetz, R. Chasan and J. A. Spudich, \textit{J. Cell Biol.} \textbf{99}, 1867 (1984).
\bibitem{81} T. Shimmen, \textit{Bot. Mag. Tokyo} \textbf{101}, 533 (1988).
\bibitem{82} K. Svoboda and S. M. Block, \textit{Cell} \textbf{77}, 773 (1994).
\bibitem{83} K. Ito, M. Ikebe, T. Kashiyama, T. Mogami, T. Kon and K. Yashimoto, \textit {J. Biol. Chem.} \textbf{282}, 19534 (2007).
\bibitem{84} N. Kamiya and M. Tazawa, \textit{Protoplasma} \textbf{46}, 394 (1956).
\bibitem{85} J. Dainty and B. Z. Ginzburg, \textit{Biochim. Biophys. Acta} \textbf{79}, 102 (1964). 
\bibitem{86} M. Tazawa and N. Kamiya, \textit{Ann. Rep. Biol. Works Fac. Sci. Osaka Univ.} \textbf{13}, 123 (1965).
\bibitem{87} A. Finkelstein, \textit{Water Movement through Lipid Bilayers, Pores, and Plasma Membranes. Theory and Reality}, John Wiley $\&$ Sons, New York, 1987.
\bibitem{88} J. I. Schroeder, \textit{J. Gen. Physiol.} \textbf{92}, 667 (1988).
\bibitem{89} R. Wayne and M. Tazawa, \textit{Protoplasma [Suppl. \textbf{2}]}, 116 (1988).
\bibitem{90} P. B. Goodwin, \textit{Planta} \textbf{157}, 124 (1983).
\bibitem{91} G. Blobel, \textit{Proc. Natl. Acad. Sci. USA} \textbf{77}, 1496 (1980).
\bibitem{92} G. E. Palade, \textit{Science} \textbf{189}, 347 (1975).
\bibitem{93} J. E. Rothman, \textit{Harvey Lectures} \textbf{86}, 65 (1992).
\bibitem{94} R. W. Schekman, \textit{Harvey Lectures} \textbf{90}, 41 (1996).
\bibitem{95} W. Haupt, \textit{Ann. Rev. Plant Physiol.} \textbf{33}, 204 (1982).
\bibitem{96} S. Takagi, E. Kamitsubo and R. Nagai, \textit{Protoplasma} \textbf{168}, 153 (1992).
\bibitem{97} K. Luby Phelps, D. L. Taylor and F. Lanni, \textit{J. Cell Biol.} \textbf{102}, 2015 (1986) .
\bibitem{98} E. Kamitsubo, M. Kikuyama and I. Kaneda, \textit{Protoplasma [Suppl. \textbf{1}]}, 10 (1988).
\bibitem{99} S. Inou\'e, in \textit{Biophysical Science-A Study Program} eds. J. L. Oncley, F. O. Schmitt, R. C. Williams, M. D. Rosenberg and R. H. Holt, John Wiley $\&$ Sons, Inc., New York, 1959, p. 402.
\bibitem{100} R. B. Nicklas, \textit{J. Cell Biol.} \textbf{97}, 542 (1983).
\bibitem{101} P. K. Hepler, \textit{J. Cell Biol.} \textbf{100}, 1363 (1985).
\bibitem{102} M. A. Sleigh, \textit{The Biology of Cilia and Flagella}, Macmillan, New York, 1962.
\bibitem{103} R. Kamiya and G. B. Witman, \textit{J. Cell Biol.} \textbf{98}, 97 (1984).
\bibitem{104} Y. Masuda, \textit{Bot. Mag. Tokyo} Special Issue \textbf{1}, 103 (1978).
\bibitem{105} J.-P. M\'etraux and L. Taiz, \textit{Plant Physiol.} \textbf{61}, 135 (1978).
\bibitem{106} A. Okamoto-Nakazato, \textit{J. Plant Res.} \textbf{115}, 309 (2002).
\bibitem{107} T. E. Proseus and J. S. Boyer, \textit{Ann. Bot.} \textbf{98}, 93 (2006).
\bibitem{108} E. M. Purcell, \textit{Amer. J. Phys.} \textbf{45}, 3 (1977).
\bibitem{109} O. Reynolds, in \textit{Papers on Mechanical and Physical Subjects} Vol. II. (1881-1900), Cambridge University Press, Cambridge, 1901, p. 51. 
\bibitem{110} N. Kamiya and K. Kuroda, \textit{Protoplasma} \textbf{50}, 144 (1958).
\bibitem{111} M. Tazawa and U. Kishimoto, \textit{Plant Cell Physiol.} \textbf{9},  361 (1968).
\bibitem{112} R. E. Williamson, \textit{J. Cell Sci.} \textbf{17}, 655 (1975).
\bibitem{113} S. Vogel, \textit{Life in Moving Fluids. The Physical Biology of Flow}, Princeton University Press, Princeton, 1981.
\bibitem{114} M. W. Denny, \textit{Air and Water. The Biology and Physics of Life's Media}, Princeton University Press, Princeton, 1993.
\bibitem{115} M. Planck, \textit{Theory of Heat}, Macmillan, New York, 1949.
\bibitem{116} Many biophysical cell biologists are familiar with microscopes based on the Doppler Effect (J. Earnshaw and M. Steer, \textit{Proc. Roy. Micro. Soc.} \textbf{14}, 108 (1979).
\bibitem{117} H. E. Ives and G. R. Stillwell, \textit{J. Opt. Soc. Am.} \textbf{28}, 215 (1938).
\bibitem{118} My treatment is not based on the common assumption that an electron is a mathematical point. Even the existence of a mathematical point is nothing more than an unproven definition used by Euclid to build his system of geometry; and the success of geometry does not prove that a mathematical point exists in reality (J. L. Synge, \textit{Science, Sense and Nonsense} W. W. Norton $\&$ Co., New York, 1951). After observing a ``point'' under the microscope, Robert Hooke (R. Hooke, \textit{Micrographia} Printed by Jo. Martyn and Ja. Allestry. London, 1665) found that the ``Point of a Needle [which] is commonly reckon'd for one\ldots if view'd with a very good Microscope… appears a broad, blunt, and very irregular end; not resembling a Cone, as is imagin'd\ldots.'' Likewise, while the idea of defining an elementary particle as a mathematical point is a good starting point in science, it too is making use of an unproven definition and better resolution of the particle in the real world may lead to a visualization of its extension. In this paper, I assume that all elementary particles in reality have extension and moving particles have a ``front'' and a ``back''. The assumption that an electron has extension requires the additional assumption that the charge is indivisible.
\bibitem{119} Since the light source, mirrors and detector in the interferometer are all in the same inertial frame, in the Michelson-Morley experiment \textit{v} = 0 and the speed of light given by eq. 4 would be the same in any and all directions.
\bibitem{120} A. A. Michelson and E. W. Morley, \textit{Amer. J. Sci.} \textbf{34},  333 (1887).
\bibitem{121} A. A. Michelson, \textit{Light Waves and their Uses}, University of Chicago Press, Chicago, 1907.
\bibitem{122} M. Planck, \textit{Scientific Autobiography and Other Papers}, Williams $\&$ Norgate, London, 1950.
\bibitem{123} A. Einstein, 1907. in \textit{The Collected Papers of Albert Einstein}. Vol. 2. English Translation, Princeton University Press, Princeton, 1989, p. 232.
\bibitem{124} S. Reinhardt, G. Saathoff, H. Buhr, L. A. Carlson, A. Wolf, D. Schwalm, S. Karpuk, C. Novotny, G. Huber, M. Zimmermann, R. Holzwarth, T. Udem, T. W. H\"ansch and G. Gwinner, \textit{Nature Physics} \textbf{3}, 861 (2007).
\bibitem{125} A. Einstein, 1917. in \textit{The World of the Atom}. eds. H. A. Boorse and L. Motz, Basic Books, New York, 1966, p. 888.
\bibitem{126} Later in this paper, the momentum of the source will be estimated by the momentum of the photons at the peak of the black body distribution curve.
\bibitem{127} A. H. Compton, \textit{Phys. Rev.} \textbf{21}, 483 (1923).
\bibitem{128} E. Feenberg and H. Primakoff, \textit{Phys. Rev.} \textbf{73}, 449 (1948).
\bibitem{129} This analysis assumes that the photon is re-emitted isotropically  \cite{125}. If the radiation is reflected or emitted at the same angle that it is absorbed, the viscous force would be twice as large.
\bibitem{130} Assuming isotropy in the center of moment frame of the photon gas, where the linear momentum coming from any direction $\hbar k$($\theta$,$\varphi$) = $\hbar k$(0,0), the total linear momentum coming from all directions is $\hbar k$ = $\int^{2\pi}_{0}$ $\int^{\pi}_{0}$ $\hbar k$($\theta$,$\varphi$) sin$\theta$ d$\theta$ d$\varphi$ = $4\pi\hbar k(0,0)$. The total linear momentum coming from all directions per unit area per unit time is $\hbar k$ $\int^{2\pi}_{0}$ $\int^{\frac{\pi}{2}}_{0}$ cos $\theta$ sin $\theta$ d$\theta$ d$\varphi$ = $\frac{\hbar k}{4}$.
\bibitem{131} W. Nernst, \textit{The New Heat Theorem. Its Foundations in Theory and Experiment}, Dover, New York, 1969.
\bibitem{132} This way of thinking also applies to neutral particles, including neutrons and neutrinos that have a magnetic moment that may form an electrical dipole that can couple to the radiation field. It would not apply to uncharged particles without a magnetic moment.
\bibitem{133} Let \textit{x} = $\frac{hc}{\lambda k \textit{T}}$. $\int$ $\frac{x^{3}}{[exp(\textit{x})-1]}$ \textit{dx} = $\frac{\pi^{4}}{15}$.
\bibitem{134} H. A. Lorentz, \textit{Nature} \textbf{113}, 608 (1924).
\bibitem{135} R. Wayne, in, R. Wayne, \textit{Light and Video Microscopy}, Elsevier Academic Press, Amsterdam, 2009, p. 277.
\bibitem{136} L. Silberstein, \textit{Phil. Mag. Ser. 6} \textbf{44}, 257 (1922).
\bibitem{137} L. Silberstein and A. P. H. Trivelli, \textit{Phil. Mag. Ser. 6} \textbf{44}, 956 (1922).
\bibitem{138} A. P. H. Trivelli and L. Richter, \textit{Phil. Mag. Ser. 6} \textbf{44}, 252 (1922).
\bibitem{139} While these geometrical cross sections appear large, using eq. (25), the geometrical cross section calculated for a 10 MeV photon would be 1.23 x 10$^{-27}$ m$^{2}$ or 0.123 barn, within the range of the experimentally determined photon cross sections. The cross section is typically a measure of the probability that any given reaction will occur and the total cross section is a measure of the probability that all possible reactions will occur. The cross sections for individual processes that make up the total cross sections vary by many orders of magnitude and may be less than, equal to or greater than the geometrical cross section. Here I assume that a charged particle in thermal equilibrium with the black body radiation field has a resonance for photons in the radiation field with every possible angular wave number and thus the probability of an electron interacting with the radiation field is unity. Consequently, the effective cross section equals the geometrical cross section.
\bibitem{140} This calculation can also be based on the fact that light radiated from an object provides that object with linear momentum antiparallel to the direction of radiation (G. N. Lewis, \textit{Phil. Mag. Ser. 6.} \textbf{16}, 705 (1908)). Semi-classically, the equivalent momentum of the photon of light is equal to \textit{m\textbf{v}}. Since the photon travels at the speed of light (c), its equivalent momentum is given by \textit{m}c. According to quantum theory, the momentum of the photon is given by $\hbar k$. By equating the classical and quantum descriptions of momentum, the equivalent mass of a photon is given by the absolute value of $\hbar$\textit{k}/c, which is equal to $\hbar\omega$/c$^{2}$. Friedrich Hasen\"ohl derived the relationship, \textit{E} $\approx m$c$^{2}$, entirely based on classical reasoning making use of Maxwell's light pressure and equating the Poynting vector to the momentum vector multiplied by c$^{2}$ (P. Lenard, \textit{Great Men of Science. A History of Scientific Progress}, G. Bell and Sons, London, 1933; W. Pauli, Theory of Relativity, Dover, New York, 1958).
\bibitem{141} R. P. Feynman, \textit{Quantum Electrodynamics}, W. A. Benjamin, New York, 1962.
\bibitem{142} R. P. Feynman, \textit{QED. The Strange Theory of Light and Matter}, Princeton University Press, Princeton, 1985.
\bibitem{143} P. Frank, \textit{Einstein. His Life and Times}, Alfred Knopf, New York, 1947.
\bibitem{144} Since the integral of force with respect to distance gives energy, integrating eq. 31 with respect to distance gives the relationship between energy the apparent mass of a charged particle at velocities close to c.
\bibitem{145} O. Reynolds, \textit{Nature} \textbf{33}, 429 (1886).
\bibitem{146} J. A. Thomson, \textit{Introduction to Science}, Henry Holt and Company, New York, 1911.
\bibitem{147} M. Planck, \textit{Verh. der Deutschen Physikalische Gesellschaft} \textbf{8}, 136 (1906).
\bibitem{148} G. N. Lewis, \textit{Science} \textbf{30}, 84 (1909).
\bibitem{149} G. N. Lewis and R. C. Tolman, \textit{Phil. Mag. Ser. 6.} \textbf{18}, 510 (1909).
\bibitem{150} R. C. Tolman, \textit{Phil. Mag. Ser. 6.} \textbf{21}, 296 (1911a).
\bibitem{151} R. C. Tolman, \textit{Phil. Mag. Ser. 6.} \textbf{22}, 458 (1911b). 
\bibitem{152} R. C. Tolman, \textit{Phil. Mag. Ser. 6.} \textbf{23}, 375 (1912).
\bibitem{153} D. Kleppner and R. J. Kolenkow, \textit{An Introduction to Mechanics}, McGraw-Hill, New York, 1973.
\bibitem{154} W. Bertozzi, \textit{Amer. J. Phys.} \textbf{32}, 551 (1964).
\bibitem{155} J. J. Thomson, \textit{The Corpuscular Theory of Matter}, Archibald Constable $\&$ Company, London, 1907.
\bibitem{156} R. A. Sunyaev and Ya. B. Zel'dovich, \textit{Ann. Rev. Astron. Astrophys.} \textbf{18}, 537 (1980).
\bibitem{157} E. M. Purcell, \textit{Electricity and Magnetism. Second Edition}, McGraw-Hill, Boston, 1985.
\bibitem{158} W. G. V. Rosser, \textit{Classical Electromagnetism via Relativity}, Plenum Press, New York (1968).
\bibitem{159} R. D. Mattuck, \textit{A Guide to Feynman Diagrams in the Many-Body Problem}, McGraw-Hill, Boston, 1967.
\end{thebibliography}
\end{document}